\documentclass[%
reprint,
 amsmath,amssymb,
 aps,
pra,
]{revtex4-1}

\usepackage{graphicx}
\usepackage{color}
\usepackage{dcolumn}
\usepackage{bm}

\begin{document}


\title{Mechanism for giant enhancement of transport induced by active fluctuations}

\author{K. Bia{\l}as}
\affiliation{Institute of Physics, University of Silesia, 41-500 Chorz{\'o}w, Poland}
\author{J. Spiechowicz}
\email{jakub.spiechowicz@us.edu.pl}
\affiliation{Institute of Physics, University of Silesia, 41-500 Chorz{\'o}w, Poland}

\begin{abstract}
Understanding the role of active fluctuations in physics is a problem \emph{in statu nascendi} appearing both as a hot topic and a major challenge. The reason for this is the fact that they are inherently non-equilibrium. This feature opens a landscape of phenomena yet to be explored that are absent in the presence of thermal fluctuations alone. Recently a paradoxical effect has been briefly communicated in which a free particle transport induced by active fluctuations in the form white Poisson shot noise can be enormously boosted when the particle is additionally subjected to a periodic potential. In this work we considerably extend the original predictions and investigate the impact of statistics of active noise on the occurrence of this effect. We construct a toy-model of the jump-relaxation process that allow us to identify different regimes of the free particle transport boost and explain their corresponding mechanisms. Moreover, we formulate and interpret the conditions for statistics of active fluctuations that are necessary for the emergence of giant enhancement of the free particle transport induced by the periodic potential. Our results are relevant not only for microscopic physical systems but also for biological ones such as e.g. living cells where fluctuations generated by metabolic activities are active by default.

\end{abstract}

\maketitle


\section{Introduction}
Active fluctuations in contrast to thermal ones are inherently non-equilibrium what implies that they are not constrained by fundamental laws of physics like the fluctuation-dissipation theorem \cite{kubo, marconi} or detailed balance symmetry \cite{cates, gnesotto} and keep the system permanently out of equilibrium even in the absence of external perturbations. Solely this feature opens a new landscape of phenomena \cite{gammaitoni1998, hanggi2009, slapik2019, metzler2014, spiechowicz2019njp} that, to a large extent, still remains a \emph{terra incognita}. Understanding of the role of active fluctuations in living matter is emerging as a hot topic and a major challenge for physics \cite{kanazawa2020}. Fluctuations generated by metabolic activities are active by default. They can be exploited by various physiological processes. For instance, biological motors like dynein and kinesin make use of such noise to enhance their directional movement along microtubules \cite{ezber, ariga}. Other manifestation include active matter harvesting energy from environment to generate a self-propulsion \cite{cates, ramaswamy, romanczuk, marchetti, olson, bechinger} or active bath such as a suspension of active colloids that surrounds a passive system \cite{bechinger, maggi, kanazawa2015, maggi2, dabelow, lee2022}.

It is commonly expected that when the free particle coupled to thermal bath and subjected to a weak constant bias is placed inside a periodic potential its velocity will be significantly reduced due to the presence of the barriers \cite{risken}. However, recently a paradoxical effect has been briefly reported in which a free particle transport induced by active fluctuations can be boosted by many orders of magnitude when the particle is additionally subjected to a periodic potential \cite{praca_w_PRE}. It is significant for understanding non-equilibrium environments such as living cells where it can explain from fundamental point of view why spatially periodic structures known as microtubules are necessary to generate effective intracellular transport. 

In the present work we considerably extend the original predictions and investigate in detail the mechanisms of this effect. In particular, we focus on the impact of statistics of active fluctuations on the occurrence of giant enhancement of transport induced by a periodic potential. In doing so we consider a broad class of probability distributions with raising level of complexity. Moreover, we construct a toy-model of the jump-relaxation process that allows us to identify different regimes of the free particle transport boost as well as understand and formulate conditions that are necessary for the emergence of this phenomenon. 

The work is organized in the following way. In Sec. II we introduce the model of a Brownian particle exposed to active fluctuations in the form of white Poisson shot noise and additionally subjected to a periodic potential. Next, we detail on different probability distributions for amplitudes of active fluctuations and establish the rescaled velocity of the particle as a main quantity of interest in this work. In Sec. III we elaborate on the giant boost of the free particle transport in a periodic potential, in particular, we analyze the impact of the spiking frequency of white Poisson shot noise on this effect. Later, in Sec. IV by resorting to a toy model of the jump relaxation process we explain the mechanisms of detected giant boost in two distinct situations of rare active fluctuations spikes and a resonance regime. Last but not least, Sec. V provides a brief summary and final conclusions. In Appendix A we calculate the mean relaxation time in the considered periodic potential.

\section{Model}
We start our investigation with a free overdamped Brownian particle described by the following simplest Langevin equation
\begin{equation}
    \Gamma \dot{x} = \sqrt{2 \Gamma k_B T}\, \xi(t),
\end{equation}
where $x$ is the particle position and the dot denotes differentiation with respect to time $t$. $\Gamma$ stands for a friction coefficient, $k_B$ is the Boltzmann constant and $T$ describes temperature of the system. Thermal fluctuations are modeled by white Gaussian noise $\xi(t)$ of zero mean $\langle \xi(t) \rangle=0$ and the correlation function $\langle \xi(t) \xi(s) \rangle = \delta(t-s)$. The particle diffuses with the celebrated Einstein free diffusion coefficient \cite{spiechowicz2023entropy}
\begin{equation}
	D = D_0 = \frac{k_B T}{\Gamma}.
\end{equation} 
Since thermal fluctuations are symmetric the average velocity of the system vanishes 
\begin{equation}
	\langle v \rangle \equiv \langle \dot{x} \rangle = 0.
\end{equation}
where $\langle \cdot \rangle$ stands for the average over its trajectories.

If additionally a constant external force $F$ is applied to the particle its dynamics read
\begin{equation}
	\Gamma \dot{x} = F + \sqrt{2 \Gamma k_B T}\, \xi(t).
     \label{eq_f}
\end{equation}
The diffusive behavior remains unchanged and it is still described by the Einstein free diffusion coefficient \mbox{$D = D_0$} \cite{spiechowicz2023entropy}. However, in such a case due to the symmetry breaking the particle attains the finite average velocity
\begin{equation}
    \langle v \rangle = v_0 = \frac{F}{\Gamma}.
\end{equation}

As the next step let us put this forced particle into a spatially periodic potential $U(x) = U(x + L)$. The dynamics of such a system is described by the analogous Langevin equation
\begin{equation}
	\Gamma \dot{x} = -U'(x) + F + \sqrt{2 \Gamma k_B T}\, \xi(t).
\end{equation}
When the particle is exposed to the periodic force $-U'(x)$ its motion is hampered and consequently the effective diffusion coefficient is typically reduced as compared to the Einstein free diffusion $D < D_0$ \cite{spiechowicz2023entropy,barkai2023}. The same observation holds true for the particle transport $\langle v \rangle \leq v_0$ \cite{risken}. However, for a critically tilting force $F = F_c$ thermal fluctuations can cooperate with the tilted periodic potential to accelerate diffusion of a particle by many orders of magnitude as compared to free diffusion $D \gg D_0$. This mechanism is known as the \textit{giant diffusion} effect \cite{reimann, lindner2016, spiechowicz2020pre, spiechowicz2021pre2} and it serves as a seed for our main problem of interest in this study. 

Specifically, we ask whether transport of a particle dwelling in a periodic potential can be greater than for a free particle, i.e. $\langle v \rangle > v_0$? As pointed out above, this scenario is ruled out for a constant force since even when $F$ is large, the particle average velocity is at most equal to the free particle velocity $\langle v \rangle = v_0$ \cite{risken}. For this reason we replace the constant bias with active nonequilibrium fluctuations $\eta(t)$ of equal mean value, i.e. $\langle \eta(t) \rangle = F$. The dynamics of such system is described by the Langevin equation
\begin{equation}
	\Gamma \dot{x} = -U'(x) + \eta(t) + \sqrt{2 \Gamma k_B T}\, \xi(t).
    \label{eq_1}
\end{equation}
The spatially periodic potential $U(x) = U(x + L)$ is assumed to be in the simplest symmetric form
\begin{equation}
	U(x) = E \sin{\left( \frac{2 \pi}{L} x \right)},
\end{equation}
where the amplitude $E$ is half of the potential barrier height. We note that since $\langle \eta(t) \rangle = F$ the average velocity reads $\langle v \rangle = v_0$ when the particle is free $U(x) = 0$.

As a model of active nonequlibrium fluctuations $\eta(t)$ we consider white Poisson shot noise \cite{hanggi1980, spiechowicz2014pre, bialas2020, praca_w_PRE}
\begin{equation}
    \eta(t)=\sum^{n(t)}_{i=1}z_i\delta (t-t_i),
\end{equation}
where $\{z_i\}$ are independent random amplitudes of $\delta$-kicks drawn from the common probability distribution $\rho(z)$. It can be characterized by its mean $\zeta=\langle z_i\rangle$, variance $\sigma^2=\langle\left( z_i-\zeta\right)^2\rangle$ and skewness $\chi=\langle\left(z_i-\zeta\right)^3\rangle/\sigma^3$, to name only a few. The arrival times $t_i$ are determined by the Poisson process, i.e. the probability for the emergence of $k$ impulses in the interval $[0,t]$ is given as 
\begin{equation}
	Pr\{ n(t) = k \} = \frac{(\lambda t)^k}{k!} e^{-\lambda t}.
\end{equation}
The parameter $\lambda$ describes the mean number of $\delta$-spikes per unit time. Active fluctuations $\eta(t)$ form white noise of a finite mean and a covariance given by
\begin{align}
 	\langle \eta(t) \rangle &= \lambda \langle z_i \rangle \\
 	\langle \eta(t)\eta(s) \rangle - \langle \eta(t) \rangle \langle \eta(s) \rangle &= \lambda \langle z_i^2 \rangle \delta(t-s).
\end{align}
For simplicity we also assume that thermal noise $\xi(t)$ is uncorrelated with nonequilibrium noise $\eta(t)$, i.e. $\langle \xi(t) \eta(s) \rangle = \langle \xi(t) \rangle \langle \eta(s) \rangle = 0$.

\subsection{Dimensionless equation}
Analysis of dimensional equations of motion may be needlessly complicated. In physics only the relation between characteristic scales of time, length and energy but not their absolute values are crucial for disentangling underlying phenomena. Transforming the equation of interest into its dimensionless form often can simplify the problem and reduce the number of parameters. Moreover, the obtained results are independent of specific experimental setup which makes them attractive from both theoretical and experimental point of view. Upon introducing the appropriate length and time unit the original Eq. (\ref{eq_1}) can be transformed into the following dimensionless form
\begin{equation}
	\dot{\hat{x}} =-\hat{U}'(\hat{x}) + \hat{\eta}(\hat{t}) + \sqrt{2D_T}\,\hat{\xi}(\hat{t}),
	\label{dimless_model}
\end{equation}
where the periodic potential reads
\begin{equation}
	\hat{U}(\hat{x}) = \varepsilon \sin{\hat{x}}
\end{equation}
and $\varepsilon$ is half of its dimensionless barrier height. The rescaled thermal noise $\hat{\xi}(\hat{t})$ and active fluctuations $\hat{\eta}(\hat{t})$ possess the same statistical properties as the corresponding dimensional ones. In particular, the mean $\langle \hat{\xi}(\hat{t}) \rangle=0$ and the correlation function $\langle \hat{\xi}(\hat{t})\hat{\xi}(\hat{s}) \rangle = \delta(\hat{t}-\hat{s})$. Similarly, $\langle \hat{\eta}(\hat{t}) \rangle = \hat{\lambda} \langle \hat{z}_i \rangle$ and $\langle \hat{\eta}(\hat{t})\hat{\eta}(\hat{s}) \rangle - \langle \hat{\eta}(\hat{t}) \rangle \langle \hat{\eta}(\hat{s}) \rangle = \hat{\lambda} \langle \hat{z}_i^2 \rangle \delta(\hat{t}-\hat{s})$. We refer the reader to Ref. \cite{praca_w_PRE} for more details of the scaling procedure. Since from here onwards only the dimensionless quantities are used, we omit the hat notation in the following sections of the paper.
\begin{figure}
    \centering
    \includegraphics[width=1.0\linewidth]{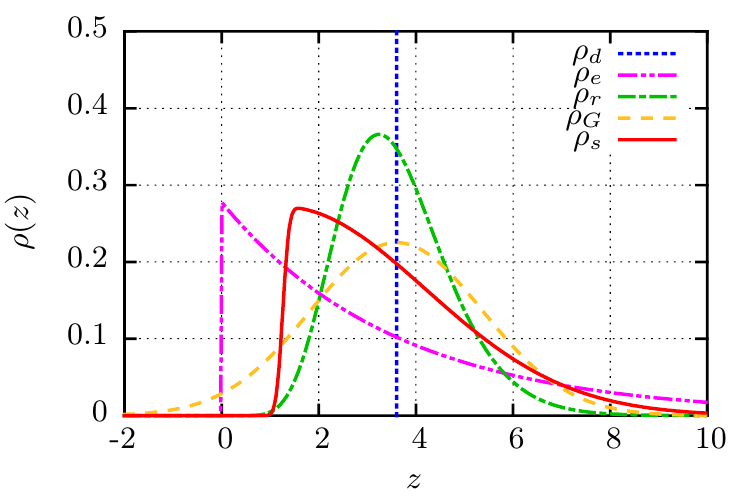}
    \caption{Comparison of the probability density functions $\rho(z)$ for amplitudes $\{z_i\}$ of $\delta$-spikes depicted for the same mean $\langle z_i \rangle = \zeta = 3.6$. The Erlang distribution $\rho_r(z)$ is illustrated for $n=10$, the variance of both Gaussian $\rho_G(z)$ and skew-normal $\rho_s(z)$ statistics is $\sigma^2=3.1$ and the skewness of the latter is $\chi=0.99$.}
    \label{fig:1}
\end{figure}
\begin{table*}[t]
\begin{ruledtabular}
\begin{tabular}{cccccc}
 Distribution&Random&Asymmetric&Non-monotonic&Bidirectional&Independent variance\\ \hline
 Deterministic $\rho_d(z)$ &no&no&no&no&no\\
 Exponential $\rho_e(z)$ &yes&yes&no&no&no\\
 Erlang $\rho_r(z)$ &yes&yes&yes&no&no\\
 Gaussian $\rho_G(z)$ &yes&no&yes&yes&yes\\
 Skew-normal $\rho_s(z)$ &yes&yes&yes&yes&yes\\
\end{tabular}
\end{ruledtabular}
\caption{\label{tab:table1} Comparison of most important properties of the probability density functions $\rho(z)$ for amplitudes of $\delta$-spikes of active fluctuations $\eta(t)$ considered in this study.}
\end{table*}

\subsection{Amplitude distributions}
In this work we aim to analyze several classes of probability distribution $\rho(z)$ for amplitudes $\{z_i\}$ of $\delta$-spikes with increasing level of complexity to explain in detail the mechanism of enhancement of the free particle transport. 
Moreover, a comprehensive discussion of a wide range of statistics would allow us to identify properties of the amplitude distribution $\rho(z)$ that are required for this effect to occur. The probability density functions taken into consideration in this study are presented in Fig. \ref{fig:1} whereas in Table \ref{tab:table1} we list their most important features for a quick reference. Our selection in principle allows to capture the following distinct physical situations: (i) an active particle in contact with a thermal bath; (ii) a passive particle surrounded by an active bath; (iii) an active particle immersed in an active bath.

\subsubsection{Deterministic active noise}
We start our analysis with the simplest case in which the amplitudes $\{z_i\}$ of $\delta$-spikes are non-random and equal, namely
\begin{equation}
    \rho_d(z)=\delta(z-\zeta),
\end{equation}
but the impulses emerge at random arrival times $\{t_i\}$ described by the Poisson process. In this study, we limit ourselves to positive bias $\langle \eta(t)\rangle > 0$ and consequently $\zeta > 0$.

\subsubsection{Exponential distribution}
Randomness of $\delta$-spikes' amplitudes is a property required to model active fluctuations. One of the simplest non-trivial probability density function is the exponential distribution. It has been widely used in countless contexts not only in physics but also in chemistry and other branches of exact and natural sciences \cite{feller1970}. In particular, in our case it can serve as a model for active fluctuations representing a self-propelling mechanism of an active particle \cite{bechinger, dabelow}. The corresponding probability density function reads
\begin{equation}
    \rho_{e}(z) =  \frac{\theta(z)}{\zeta} \mbox{exp}\left(-\frac{z}{\zeta}\right).
    \label{pdf_exp}
\end{equation}
It means that at random instants of time $t_i$ the particle is exposed to $\delta$-spikes of random non-negative amplitudes $z_i \ge 0$. Such model of active fluctuations can describe e.g. the stochastic release of energy in chemical reactions such as ATP hydrolysis \cite{chowdhury2013}. The distribution $\rho_e(z)$ is a monotonically decreasing function, meaning that the probability of $\delta$-spikes with larger amplitudes is smaller than those with the smaller ones. Moreover, the variance of exponential distribution $\sigma^2_e=\zeta^2$ is dependent on its mean. The skewness of exponential distribution is $\chi=2$ so it is also highly asymmetric.

\subsubsection{Erlang distribution}
Modeling of self-propelling mechanisms of active particles in terms of white Poisson noise $\eta(t)$ with exponentially distributed amplitudes $z_i$ suffers from one drawback. Small $\delta$-spikes are the most likely ones which is not necessarily the case. Therefore a non-monotonic probability density function $\rho(z)$ must be considered. The generalization of the exponential statistics called the Erlang distribution \cite{johnson} fulfills this condition and reads
\begin{equation}
    \rho_{r}(z) = \frac{\theta(z) z^{n-1}}{{(\zeta/n)}^n(n-1)!}  \mbox{exp}\left (-\frac{n z}{\zeta} \right),
    \label{pdf_erlang}
\end{equation}
where $n \in \mathbb{N}$. Sum of $n$ exponentially distributed random variables drawn from the same distribution with mean $\mu=\zeta/n$ follows the Erlang distribution. While its variance is still related to its mean $\sigma^2_r= \zeta^2/n$ the Erlang statistics is non-monotonic and for $n>1$ it possesses maximum at $z=(n-1)\zeta/n$. It is less asymmetric than the exponential distribution as its skewness reads $\chi=2/\sqrt{n}$.

\subsubsection{Gaussian distribution}
In all previous cases the amplitudes are non-negative $z_i \ge 0$. However, active fluctuations $\eta(t)$ may represent not only self-propelling mechanisms but also impact of an active bath such as suspension of active microswimmers on a passive or active particle \cite{kanazawa2020, dabelow, lee2022}. In this case $\eta(t)$ models random collisions with the active environment that can result in both positive and negative $\delta$-spikes. To take into account this scenario one has to consider a bidirectional amplitude distribution $\rho(z)$ allowing impulses in both directions. As one of the most obvious examples we pick the Gaussian distribution
\begin{equation}
    \rho_G(z)=\frac{1}{\sqrt{2\pi \sigma_G^2}} \exp{\left(-\frac{(z-\zeta)^2}{2\sigma^2_G}\right)}.
\end{equation}
Like the previous one this distribution renders a non-monotonic function. However, unlike the Erlang statistics, here the variance $\sigma_G^2$ is a parameter independent of its mean $\zeta$. Moreover, Gaussian distribution is symmetric around $\zeta$ and its skewness vanishes $\chi=0$.

\subsubsection{Skew-normal distribution}
In the most general case active fluctuations may have more than one physical origin. An example is an active self-propelling particle surrounded by an active bath. The corresponding probability density $\rho(z)$ for amplitude of $\delta$-spikes obviously should be bidirectional to take into account collisions supplying and taking energy from the system. However, to describe the balance between the influence of a self-propelling mechanism and an active bath an additional parameter is needed. The amplitude statistics $\rho(z)$ in which mean, variance and asymmetry can be independently varied serves as a good candidate to capture the most complex origin of active fluctuations. Consequently, the last distribution considered in this work is a skew-normal distribution, which is a generalization of the Gaussian distribution to non-zero asymmetry \cite{azz,rijal2022,bailey2021}
\begin{equation}
    \rho_s(z) = \frac{2}{\sqrt{2\pi \omega^2}}e^{-\frac{(z-\mu)^2}{2\omega^2}} \int_{-\infty}^{\alpha[(z-\mu)/\omega]} ds \, \frac{1}{2\pi}e^{-\frac{s^2}{2}},
\end{equation}
where $\mu$ is location, $\omega$ scale and $\alpha$ shape parameter. These quantities are defined in terms of statistical moments of the distribution, i.e. mean $\zeta$, variance $\sigma^2$ and skewness $\chi$ \cite{generacja,generacja2}:
\begin{subequations}
\begin{align}
\begin{split}
\alpha&=\frac{\delta}{\sqrt{1-\delta^2}}, \\
\end{split}\\
\begin{split}
\omega&=\sqrt{\frac{\sigma^2}{1- 2\delta^2/\pi}}, \\
\end{split}\\
\begin{split}
\mu&=\zeta-\delta\sqrt{\frac{2\sigma^2}{\pi(1-2\delta^2/\pi)}},\\
\end{split}
\end{align}
\label{eq_S_def}
\end{subequations}
where $\delta$ reads
\begin{equation}
    \delta=\text{sgn}(\chi)\sqrt{\frac{|\chi|^{2/3}}{(2/\pi)\{[(4-\pi)/2]^{2/3}+|\chi|^{2/3}\}}}.
    \label{eq_S_delta}
\end{equation}
We note that the skew-normal distribution $\rho_s(z)$ is non-monotonic, bidirectional, asymmetric and possesses variance which is independent of the mean.

\subsection{Quantities of interest}
The most basic quantity characterizing the directed transport of the considered Brownian particle is the average velocity
\begin{equation}
    \langle v \rangle = \lim_{t\to \infty} \frac{\langle x(t) \rangle - \langle x(0) \rangle}{t} = \lim_{t\to \infty} \frac{\langle x(t) \rangle}{t}.
    \label{av_v}
\end{equation}
where $\langle \cdot \rangle$ stands for the average over ensemble of thermal and active noise realizations. In this study we investigate how the free particle transport is modified when it is subjected to a periodic potential. For this reason, our main quantity of interest is the rescaled velocity $\langle v \rangle/v_0$, where
\begin{equation}
    v_0 = \langle \eta(t) \rangle = \lambda \langle z_i \rangle = \lambda \zeta
\end{equation}
is the average velocity of a free particle exposed to active fluctuations $\eta(t)$.

\begin{figure}[t]
    \centering
    \includegraphics[width=1.0\linewidth]{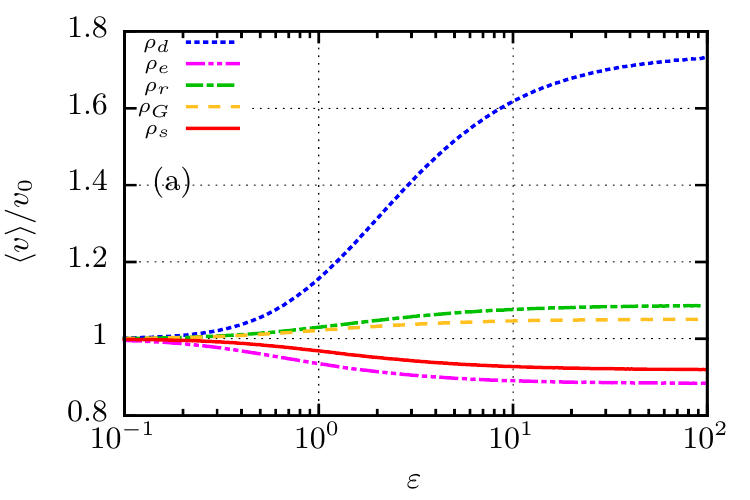}
    \includegraphics[width=1.0\linewidth]{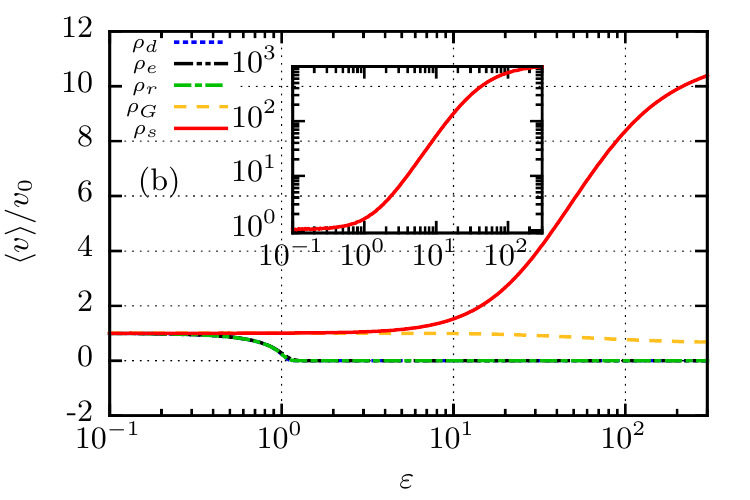}
    \caption{The rescaled average velocity $\langle v\rangle/v_0$ versus the barrier height $\varepsilon$ of the periodic potential $U(x)$ for different active fluctuations $\eta(t)$ amplitude distributions $\rho(z)$. In both panels $\langle \eta(t) \rangle = v_0 = 1$, however in (a) $\zeta=3.6$, $\lambda=1/3.6$ whereas in (b) $\zeta=1/30$ and $\lambda=30$. In the inset of panel (b) we show giant boost of the free particle transport which occurs for small $v_0 = 0.01$. Other parameters are: in the Erlang distribution $\rho_r(z)$ $n = 10$, the variance of both Gaussian $\rho_G(z)$ and skew-normal $\rho_s(z)$ statistics is $\sigma^2 = 3.1$ and the skewness of the latter is $\chi=0.99$. Thermal noise intensity is set to $D_T=0.01$.}
    \label{fig:2}
\end{figure}
\section{Giant boost of \\the free particle transport}
Unfortunately the general Fokker-Planck-Kolmogorov-Feller integro-differential equation corresponding to Eq. (\ref{dimless_model}) cannot be solved analytically in a closed way \cite{hanggi1980}. The results in literature have been attained only for some selected special cases \cite{luczka1995,kanazawa2015,talbot2011}. For this reason we performed precise computer simulations using CUDA environment on modern desktop graphics processing units (GPUs) \cite{spiechowicz2015cpc}. This approach allowed us to simulate $2^{16}$ system trajectories at once and accelerated the computation by several orders of magnitude as compared to standard methods. The ensemble averaging was performed over thermal and active noise realizations as well as over initial condition $x(0)$ distributed uniformly over the spatial period $L = 2\pi$ of the potential $U(x)$.

We start our investigation with the dependence of the rescaled velocity $\langle v\rangle/v_0$ on the barrier height $\varepsilon$ of the periodic potential $U(x)$ for different active fluctuations $\eta(t)$ amplitude distributions $\rho(z)$. This characteristic is depicted in Fig. \ref{fig:2}. In both panels we fix the statistical bias or equivalently the free transport velocity $\langle \eta(t) \rangle = v_0 = 1$. However, we distinguish two cases. In panel (a) the mean amplitude $\langle z_i \rangle = \zeta$ of active fluctuations is larger than the distance $\zeta_c = L/2= \pi$ from minimum to maximum of the potential $U(x)$, i.e. $\zeta > \zeta_c$. It means that on average when the $\delta$-spike arrives the particle is taken over the potential barrier. On the other hand, in panel (b), $\zeta < \zeta_c$ and statistically the $\delta$-kick does not transport the particle over the potential barrier.

Unless stated otherwise from now on we use the following parameters: for the Erlang distribution $\rho_r(z)$ $n = 10$. Both the Gaussian $\rho_G(z)$ and skew-normal $\rho_s(z)$ statistics have variance $\sigma^2=3.1$. Moreover, the latter possesses the asymmetry $\chi = 0.99$. Thermal noise intensity is set to $D_T = 0.01$. 

In Fig. \ref{fig:2} we observe in accordance to common intuition that when the potential barrier vanishes $\varepsilon \to 0$ the rescaled velocity $\langle v \rangle/v_0 \to 1$, i.e. the directed transport is the same as for the free particle, regardless of the value of mean amplitude $\zeta$, c.f. panel (a) and (b). When the barrier $\varepsilon$ increases the rescaled velocity $\langle v \rangle/v_0$ starts to diverge from its value characteristic for the free particle. In particular, if the mean amplitude $\zeta > \zeta_c$, see panel (a), the rescaled velocity $\langle v \rangle/v_0$ tends to different constant values depending on active fluctuations amplitude statistics $\rho(z)$. Both increase as well as  decrease of the free particle transport is possible. The largest enhancement is observed for the deterministic distribution $\rho_d(z)$ followed by much smaller one detected for the Erlang $\rho_r(z)$ and Gaussian $\rho_G(z)$ statistics. On the other hand, active fluctuations $\eta(t)$ with $\delta$-spikes distributed according to the exponential $\rho_e(z)$ and skew-normal $\rho_s(z)$ density decrease the transport when the free particle is additionally subjected to the periodic force $U(x)$. 

The situation is radically different in panel (b) where the mean amplitude is smaller than half of the spatial period of potential $\zeta < \zeta_c$. Then only active fluctuations $\eta(t)$ with $\delta$-spikes distributed according to the skew-normal $\rho_s(z)$ density lead to amplification of the free particle transport if the latter is additionally exposed to the periodic potential $U(x)$. Note that the enhancement of rescaled velocity $\langle v \rangle/v_0$ is almost an order of magnitude greater than for the deterministic distribution $\rho_d(z)$ in panel (a). Moreover, in the inset we show that if initially the free particle transport induced by active fluctuations $\eta(t)$ is smaller, e.g. $v_0 = 0.01$, the boost in rescaled velocity when the particle dwells in the periodic potential $U(x)$ can be enormous $\langle v \rangle/v_0 = 1000 \gg 1$ \cite{praca_w_PRE}. The reader should also note that for the skew-normal distribution $\rho_s(z)$ the velocity $\langle v \rangle/v_0$ reaches plateau when the potential barrier $\varepsilon \to \infty$. Such giant transport boost is significant for understanding nonequilibrium environments such as living cells where it can explain from fundamental point of view why spatially periodic structures known as microtubules are necessary to generate impressively effective intracellular transport. The goal of the present paper is to explain the mechanism of this effect in depth.
\begin{figure}[t]
    \centering
    \includegraphics[width=1.0\linewidth]{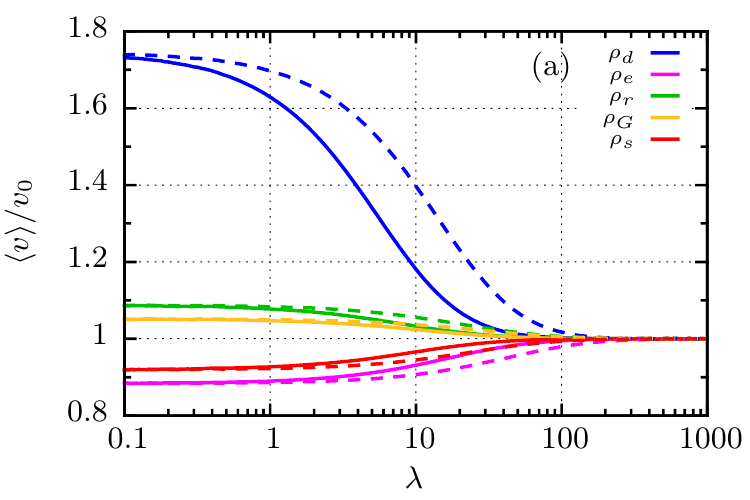}
    \includegraphics[width=1.0\linewidth]{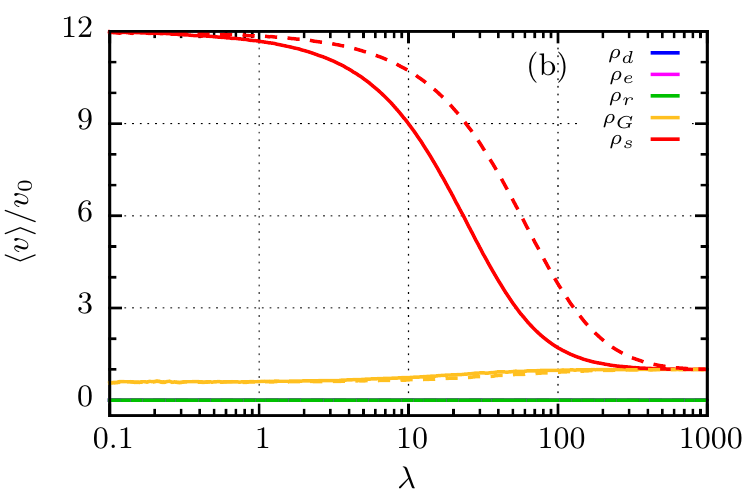}
    \caption{The rescaled average velocity $\langle v\rangle/v_0$ versus the mean spiking rate $\lambda$ depicted for different amplitude statistics $\rho(z)$ with fixed mean $\zeta = const.$. Solid lines correspond to the barrier height $\varepsilon=40$ of the periodic potential $U(x)$ and dashed ones to $\varepsilon=100$. In panel (a) results for $\zeta = 3.6 > \zeta_c$ are shown while in panel (b) for $\zeta = 1/30 < \zeta_c$. Other parameters are the same as in Fig. \ref{fig:2}.}
    \label{fig:3}
\end{figure}
\begin{figure}[t]
    \centering
    \includegraphics[width=1.0\linewidth]{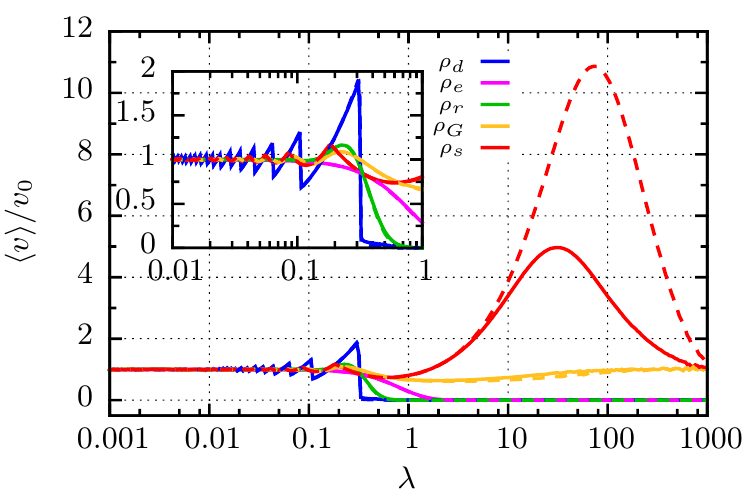}
    \caption{The rescaled average velocity $\langle v\rangle/v_0$ versus the mean spiking rate $\lambda$ depicted for different amplitude statistics $\rho(z)$ with fixed mean bias $\langle \eta(t) \rangle = v_0 = 1$. Solid lines correspond to the barrier height $\varepsilon=40$ of the periodic potential $U(x)$ and dashed ones to $\varepsilon=100$. Other parameters are the same as in Fig. \ref{fig:2}.}
    \label{fig:4}
\end{figure}

\subsection{Impact of the spiking frequency $\lambda$}
In doing so we now focus on the impact of the mean spiking frequency $\lambda$ on the boost of free particle transport driven by active fluctuations $\eta(t)$ when the system is additionally subjected to the periodic potential $U(x)$. Since the average velocity of free particle $v_0 = \langle \eta(t) \rangle = \lambda \zeta$ we distinguish two cases where the mean spiking rate $\lambda$ is varied. In the first one the mean amplitude of active fluctuations $\delta$-kicks is fixed $\zeta = const.$. However, it implies that when the spiking rate $\lambda$ is changed the free particle transport $v_0 = \lambda \zeta$ varies as well. In the second case the latter characteristic is constant $v_0 = const.$ so the mean amplitude $\zeta$ changes together with the spiking rate $\lambda$. It seems that it is more adequate option when discussing the free particle transport boost, however, the first scaling provides important additional insights into the preconditions for this effect to emerge and therefore we start our discussion with it.
\begin{figure*}[t]
    \centering
    \includegraphics[width=0.49\linewidth]{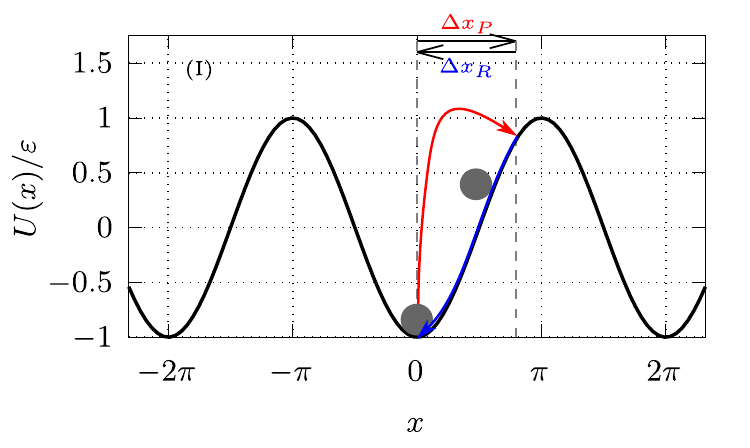}
    \includegraphics[width=0.49\linewidth]{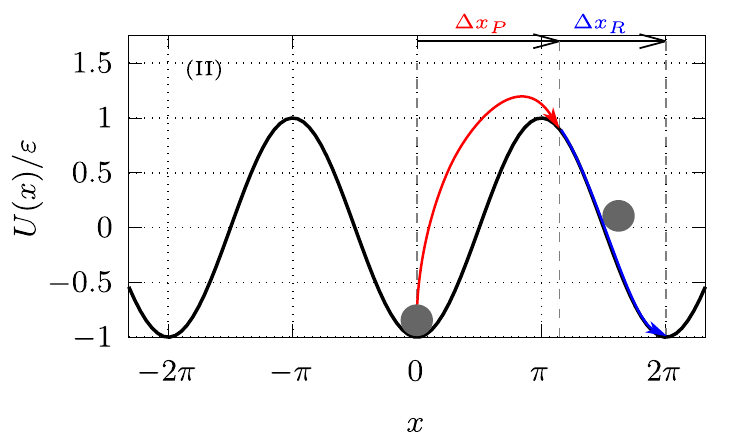}
    \includegraphics[width=0.49\linewidth]{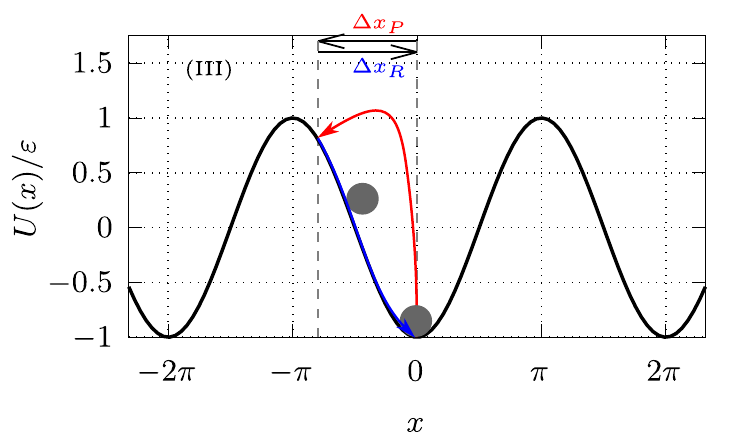}
    \includegraphics[width=0.49\linewidth]{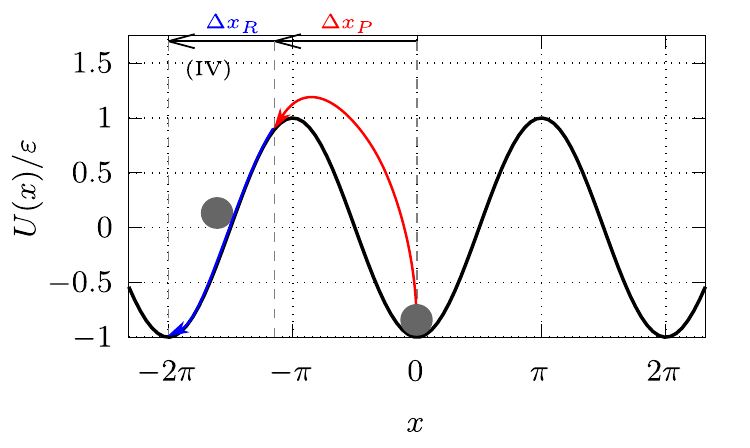}
    \caption{Schematic representation of the four (I)-(IV) elementary realizations of the particle jump-relaxation process as a phenomenological description of the simplified dynamics with neglected thermal noise contribution. The bottom ones are possible only for the bidirectional amplitude distributions $\rho(z)$ allowing for both positive and negative $\delta$-spikes, i.e. for the Gaussian $\rho_G(z)$ and skew-normal $\rho_s(z)$ statistics.}
    \label{fig:5}
\end{figure*}

\subsubsection{Fixed mean amplitude $\zeta = const.$}
In Fig. \ref{fig:3} we present how the rescaled velocity $\langle v\rangle/v_0$ changes when the mean spiking rate $\lambda$ is varied for the fixed mean amplitude $\zeta$ of $\delta$-impulses and different statistics $\rho(z)$. Moreover, solid lines correspond to the barrier height $\varepsilon = 40$ whereas the dashed ones to $\varepsilon = 100$. In the same way as before, in panel (a) the mean amplitude is supercritical $\zeta = 3.6 > \zeta_c$ while in (b) it is subcritical $\zeta = 1/30 < \zeta_c$. Note that the free particle transport $v_0 = \lambda \zeta$ changes here together with the spiking rate $\lambda$. There are a number of important observations that comes from the inspection of Fig \ref{fig:3}. Firstly, the largest free particle boost occurs in the limit of rare $\delta$-spikes $\lambda \to 0$. However, it is gigantic $\langle v \rangle/v_0 \gg 1$ only for the skew-normal $\rho_s(z)$ amplitude statistics provided that its mean $\zeta$ is significantly smaller than the distance between the minimum and maximum of the periodic potential $\zeta < \zeta_c$. The free particle transport boost observed in the limit of rare $\delta$-spikes $\lambda \to 0$ is robust with respect to variation of the potential barrier height $\varepsilon$. Secondly, when the the $\delta$-impulses are very frequent, i.e. for $\lambda \to \infty$, there is no free transport enhancement regardless of the amplitude statistics $\rho(z)$ and its mean amplitude $\zeta$. Thirdly, when the spiking frequency $\lambda$ is moderate the potential barrier height $\varepsilon$ modifies the rescaled velocity of the particle. In particular, when $\varepsilon$ grows the characteristics $\langle v \rangle/v_0$ is shifted towards larger $\lambda$ and consequently transport boost starts to be detected for greater frequencies $\lambda$.

\subsubsection{Fixed mean bias $\langle \eta(t) \rangle = v_0 = const.$}
Behavior of the rescaled velocity $\langle v\rangle/v_0$ is radically different when the condition of fixed mean $\langle \eta(t) \rangle = v_0 = const.$ is imposed. It means that the mean amplitude $\zeta$ of $\delta$-spikes changes together with the spiking rate $\lambda$. We present this case in Fig. \ref{fig:4}. Now the rescaled velocity $\langle v \rangle/v_0$ is no longer a monotonic function of the spiking rate $\lambda$. When $\delta$-kicks are scarce $\lambda \ll 1$ the transport is roughly speaking equivalent to motion of the free particle $\langle v \rangle/v_0 = 1$. However, in this parameter regime characteristic oscillatory behavior is detected as $\lambda$ grows, see the inset. 
On the other hand, if the spiking rate is very large $\lambda \to \infty$ and consequently small $\delta$-impulses are very frequent, the transport ceases to exist $\langle v \rangle/v_0 = 0$ except of two bidirectional amplitude distributions with variance $\sigma^2$ independent of mean $\zeta$, i.e. Gaussian $\rho_G(z)$ and skew-normal $\rho_s(z)$ statistics for which it is equal to the velocity of free particle $\langle v \rangle/v_0 = 1$. The most important finding is that only for the skew normal distribution $\rho_s(z)$ of amplitudes as the spiking rate $\lambda$ of $\delta$-kicks grows there exist a well pronounced maximum in the studied characteristics. The optimal $\lambda$ for which the free particle transport boost is maximal depends on the potential barrier height $\varepsilon$ and increases when $U(x)$ is steeper. Moreover, the magnitude of the free transport amplification grows together with $\varepsilon$ and can be enormous $\langle v \rangle/v_0 \gg 1$.
\begin{table*}[t]
\begin{ruledtabular}
\begin{tabular}{ccccc}
 Total displacement $\Delta x$ &Scenario I&Scenario II&Scenario III&Scenario IV\\ \hline
 Free particle $\Delta x^F$ & $\Delta x_P$ & $\Delta x_P$ & $-| \Delta x_P |$ & $-| \Delta x_P |$\\
 Particle in a periodic potential $\Delta x^P$ & $\Delta x_P - |\Delta x_R|$ & $\Delta x_P + | \Delta x_R|$ & $-|\Delta x_P| + |\Delta x_R|$ & $- |\Delta x_P| - |\Delta x_R|$ \\ \hline
 Net difference $\Delta x^P - \Delta x^F$ & $-|\Delta x_R|$ & $|\Delta x_R|$ & $|\Delta x_R|$ & $-|\Delta x_R|$ \\
\end{tabular}
\end{ruledtabular}
\caption{\label{tab:table2} Analysis of the total particle displacement $\Delta x$ for the jump process (free particle) and jump-relaxation process (particle in a periodic potential) driven by active fluctuations $\eta(t)$ in each scenario presented in Fig. \ref{fig:5}. The net difference between the particle in a periodic potential and free one tells that for positive statistical bias $\langle \eta(t) \rangle = v_0 > 0$ scenario (II) and (III) boost the transport whereas (I) and (IV) decrease it.}
\end{table*}

\section{The mechanism}
In this section we want to explain in detail what is the mechanism standing behind the giant enhancement of free particle transport induced by active fluctuations when the latter is additionally subjected to the periodic potential. For this purpose let us first note that in the presented parameter regime thermal noise intensity $D_T = 0.01$ is significantly smaller than the potential barrier height $\varepsilon$ as well as mean statistical bias of active fluctuations $\langle \eta(t) \rangle = v_0 = 1$. Therefore without loss of generality we neglect thermal noise $\xi(t)$ and discuss a phenomenological description of the simplified dynamics in terms of the jump-relaxation process. While this picture is strictly valid only in the limit of rare spikes $\lambda \to 0$ it will allow us to gain a physical intuition  on which we will later base.

\subsection{Jump-relaxation process}
In Fig. \ref{fig:5} we present the schematic representation of the four elementary realizations of the particle jump-relaxation process. Initially the particle resides at the potential minimum. When a $\delta$-spike arrives it is transported over the distance $\Delta x_P = z_i$ and then during the relaxation towards the nearest potential minimum it covers the interval $\Delta x_R$. The total particle displacement in the jump-relaxation process reads $\Delta x = \Delta x_P + \Delta x_R$. Depending on the magnitude of amplitude $z_i$ the particle may or may not overcome the potential barrier $\varepsilon$. If $0 < z_i < \zeta_c = \pi$ (scenario (I)) the distance is $\Delta x = 0$. For $\zeta_c < z_i < L = 2 \pi$ (scenario (II)) the total displacement is $\Delta x = L = 2\pi$. If $-\zeta_c < z_i < 0$ (scenario (III)) $\Delta x = 0$ whereas when $-L < z_i < -\zeta_c$ (scenario (IV)) the distance is $\Delta x = -L = -2\pi$. We limit our consideration to the interval $[-L,L]$ since further displacement in either direction can be effectively reduced to the above four classes of realizations. 

When the periodic potential is switched off $U(x) = 0$ the relaxation part $\Delta x_R$ in the total distance vanishes $\Delta x = \Delta x_P$ and consequently the free particle motion can be represented as a jump process alone. Analysis of the full displacement $\Delta x$ for the jump and jump-relaxation process presented in Table \ref{tab:table2} reveals that for positive statistical bias $\langle \eta(t) \rangle = v_0 > 0$ the transport is boosted when the particle is subjected to a periodic potential and scenario (II) or (III) take place. On the other hand the movement is slowed down in scenario (I) and (IV). We note that the case (III) and (IV) is possible only for bidirectional amplitude statistics, namely the Gaussian $\rho_G(z)$ and skew-normal distribution $\rho_s(z)$, that allow for both positive $z_i > 0$ and negative $z_i < 0$ $\delta$-spikes.
\begin{figure*}[t]
    \centering
    \includegraphics[width=0.49\linewidth]{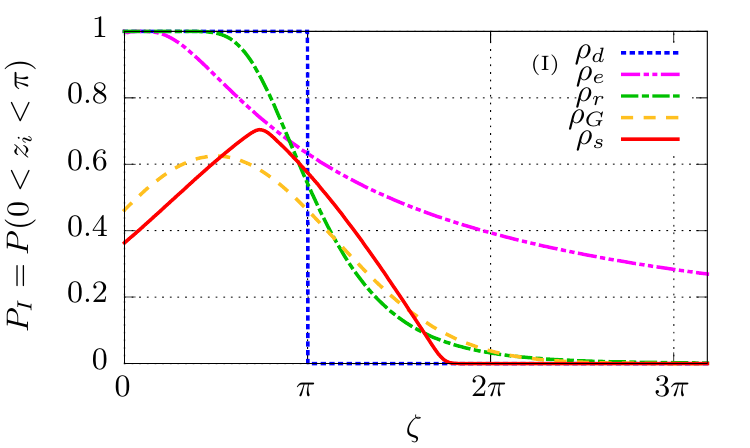}
    \includegraphics[width=0.49\linewidth]{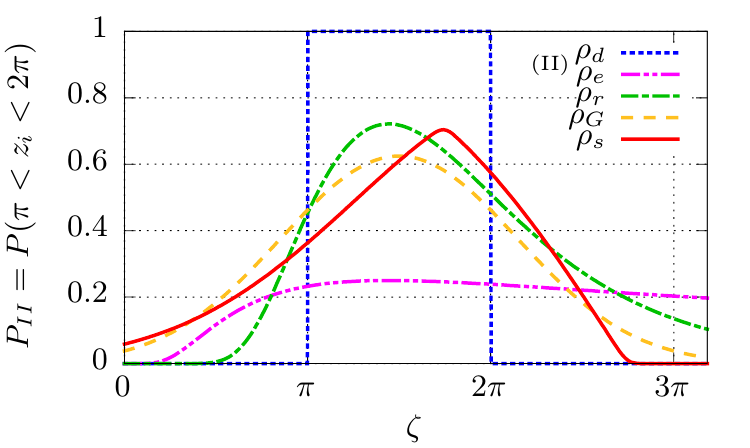}
    \includegraphics[width=0.49\linewidth]{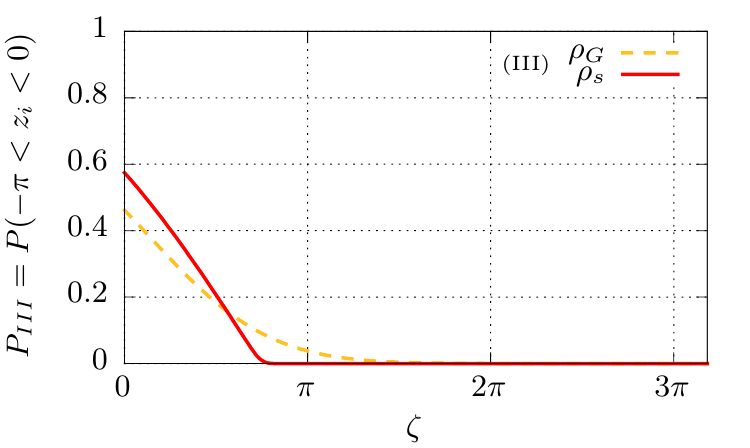}
    \includegraphics[width=0.49\linewidth]{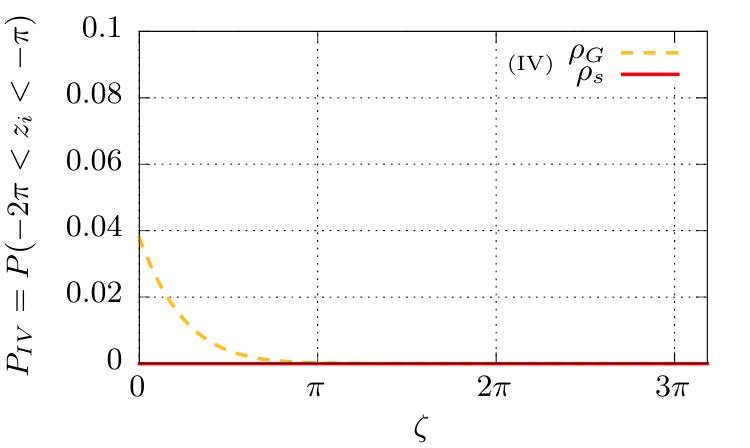}
    \caption{The probabilities $P_I$, $P_{II}$, $P_{III}$ and $P_{IV}$ for occurrence of the corresponding scenario in the jump-relaxation process, see Fig. \ref{fig:5}, as a function of the mean amplitude $\zeta$ for different variants of the active fluctuations $\delta$-spike statistics $\rho(z)$. The parameters are: for the Erlang distribution $\rho_r(z)$ $n = 10$, the variance of both the Gaussian $\rho_G(z)$ and skew-normal statistics $\rho_s(z)$ $\sigma^2 = 3.1$ and the skewness of the latter is $\chi = 0.99$.}
    \label{fig:6}
\end{figure*}

Knowing solely the amplitude distribution $\rho(z)$ it is possible to determine the probability of each scenario in a single act of $\delta$-spike action. For instance, the probability for occurrence of the second case reads
\begin{equation}
	P_{II} = Pr\{ \zeta_c = \pi < z_i < L = 2\pi \} = \int_{\zeta_c}^{L} \rho(z) \,dz.
\end{equation}
The analogous expressions for the remaining scenarios can be written down by changing the bottom and upper integration limit accordingly. In Fig. \ref{fig:6} we present the four probabilities $P_I$, $P_{II}$, $P_{III}$ and $P_{IV}$ as a function of the mean amplitude $\zeta$ for different variants of the active fluctuations $\delta$-spike statistics $\rho(z)$. Since for the deterministic distribution $\rho_d(z)$ there is only one amplitude $z_i = \zeta$ the probability $P_I$ or $P_{II}$ equals either $0$ or $1$. The exponential statistics $\rho_e(z)$ is monotonically decreasing function and consequently for this case the probability $P_I$ is always larger than $P_{II}$. The Erlang distribution $\rho_r(z)$ does not possess this property. However, both of these statistics have variance determined by the amplitude mean $\sigma^2 \propto \zeta^2$ and therefore for small $\zeta$ the probability $P_{I}$ is close to one. We note that for non-monotonic distributions the most probable scenario changes together with the mean amplitude $\zeta$. In particular, for the case of the Gaussian $\rho_G(z)$ and skew-normal $\rho_s(z)$ statistics the probabilities presented in Fig. \ref{fig:6} are only shifted as $\zeta$ grows. Last but not least, $P_{III}$ and $P_{IV}$ are non-zero only for the Gaussian $\rho_G(z)$ and skew-normal $\rho_s(z)$ distributions.

What is crucial for the boost of free particle transport driven by active fluctuations in a periodic potential is however not a magnitude of the individual probabilities $P_{I}$, $P_{II}$, $P_{III}$ and $P_{IV}$ but the balance between them. As we revealed in the analyzed case the second and the third scenario boost the free particle transport whereas the first and the fourth decrease it. Therefore we now want to analyze the difference
\begin{equation}
	\Delta P = P_+ - P_- = (P_{II} + P_{III}) - (P_{I} + P_{IV}).
\end{equation}
This quantity is depicted in Fig. \ref{fig:7} versus the mean amplitude $\zeta$ of different $\delta$-spikes statistics $\rho(z)$. A number of very important conclusions for our further analysis come from the inspection of this panel. First and foremost, for vanishing mean amplitude $\zeta \to 0$ the balance $\Delta P$ is positive -- indicating potential for the free particle transport boost -- only for the skew-normal statistics $\rho_s(z)$. On the other hand, the difference $\Delta P$ is always negative only for the exponential distribution $\rho_e(z)$ due to its monotonically decreasing form. As typically $\Delta P < 0$ when $\zeta < \zeta_c$ the free particle transport is expected to be hampered in this regime when the periodic potential is switched on. In contrast, if $\zeta > \zeta_c$ then usually the difference $\Delta P > 0$ and the transport boost can emerge.

The average velocity of the particle in the jump-relaxation process by definition can be expressed as
\begin{equation}
	\langle v \rangle = \frac{\Delta x}{\tau} = \frac{\Delta x_P + \Delta x_R}{\tau_P + \tau_R}
\end{equation}
where $\tau_P$ is a waiting time for the arrival of $\delta$-spike of active fluctuations $\eta(t)$ and $\tau_R$ is a time interval in which the particle covers the distance $\Delta x_R$ towards the potential minimum until the next $\delta$-spike emerges and the sequence is repeated, see Fig. \ref{fig:5}. 

\subsection{Regimes of the free particle transport boost}
The mean waiting time expresses the characteristic time scale of active fluctuations $\eta(t)$ and it is determined by the inverse of their spiking frequency $\langle \tau_P \rangle = 1/\lambda$. On the other hand, the average relaxation time describes the characteristic time scale of the periodic potential $U(x)$ and is proportional to the inverse of its barrier height $\langle \tau_R \rangle \propto 1/\varepsilon$, see the Appendix. The relation between these two characteristic time scales of the problem is crucial for understanding of the free particle transport boost in the periodic potential, in particular the giant one. There are three distinct regimes: (i) the rare spikes $\langle \tau_P \rangle \gg \langle \tau_R \rangle$, (ii) the frequent impulses $\langle \tau_P \rangle \ll \langle \tau_R \rangle$ and (iii) the resonance $\langle \tau_P \rangle \approx \langle \tau_R \rangle$ which we now analyze one by one.

\subsubsection{Rare spikes $\langle \tau_P \rangle \gg \langle \tau_R \rangle$}
Our assumption in the jump-relaxation process that initially the particle resides in the potential minimum is strictly satisfied only when the average waiting time between the successive $\delta$-spikes is much larger than the typical relaxation time $\langle \tau_P \rangle \gg \langle \tau_R \rangle$. In such a case the latter time scale can be neglected and the average velocity of the particle reads
\begin{equation}
	\langle v \rangle = \frac{\langle \Delta x \rangle}{\langle \tau_P \rangle} = \lambda (\langle \Delta x_P \rangle + \langle \Delta x_R \rangle) = \lambda (\zeta + \langle \Delta x_R \rangle).
\end{equation} 
Consequently the rescaled velocity is
\begin{equation}
    \frac{\langle v\rangle}{v_0}=\frac{\lambda( \zeta+\langle\Delta x_R\rangle)}{\lambda \zeta}=1+\frac{\langle \Delta x_R\rangle}{\zeta}.
    \label{eq:R_zeta}
\end{equation}
It means that to optimize free transport boost the average distance $\langle \Delta x_R \rangle$ should be maximized with the mean amplitude $\zeta$ simultaneously being minimized. Since for the studied periodic potential $U(x)$ the relaxation is limited by the distance between its minimum and maximum $\langle \Delta x_R \rangle \le \zeta_c = L/2$ we get the upper bound for the enhancement
\begin{equation}
	\frac{\langle v \rangle}{v_0} \le 1 + \frac{\zeta_c}{\zeta} = 1 + \frac{L}{2\zeta}.
	\label{eq:bound}
\end{equation}
This result tells that in the regime of rare spikes $\langle \tau_P \rangle \gg \langle \tau_R \rangle$ the free particle transport boost emerges since active fluctuations $\eta(t)$ allows the particle to exploit the spatial periodicity $L$ of the potential $U(x)$. This upper bound can be reached for the deterministic amplitude statistics $\rho_d(z)$ when $\zeta \to \zeta_c^+$ and then according to Eq. (\ref{eq:bound}) the maximal rescaled velocity reads $\lim_{\zeta \to \zeta_c^+} \langle v \rangle/v_0 = 2$.
\begin{figure}[t]
    \centering
    \includegraphics[width=1.0\linewidth]{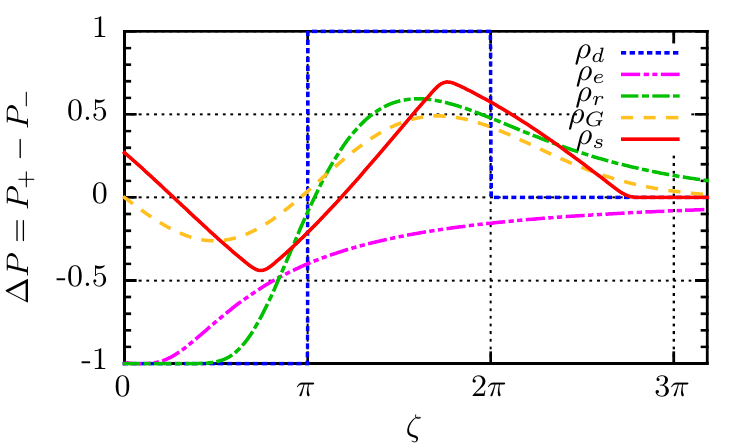}
    \caption{The difference $\Delta P = P_+ - P_- = (P_{II} + P_{III}) - (P_{I} + P_{IV})$ between the probabilities $P_+$ and $P_-$ for the occurrence of the jump-relaxation processes that are constructive and destructive for the free particle transport boost in a periodic potential, respectively. $\Delta P$ is shown versus the mean amplitude $\zeta$ of different $\delta$-spikes statistics $\rho(z)$. The parameters are the same as in Fig. \ref{fig:6}.}
    \label{fig:7}
\end{figure}

Let us now come back to Fig. \ref{fig:3} and reinterpret the results shown there in terms of the above discussion. The limit of rare spikes $\langle \tau_P \rangle \gg \langle \tau_R \rangle$ is realized when the frequency $\lambda \to 0$. In both panels of Fig. \ref{fig:3} in this regime the rescaled velocity $\langle v \rangle/v_0$ attains its plateau. In plot (a) for which $\zeta = 3.6 > \zeta_c$ the free particle transport boost $\langle v \rangle/v_0 > 1$ occurs for the deterministic $\rho_d(z)$, the Erlang $\rho_r(z)$ as well as the Gaussian  $\rho_G(z)$ distributions whereas for the skew-normal $\rho_s(z)$ and exponential $\rho_e(z)$ statistics it is hampered $\langle v \rangle/v_0 < 1$. These facts can be predicted a priori from the inspection of Fig. \ref{fig:7} where for $\zeta = 3.6$ the difference $\Delta P$ is positive for $\rho_d(z)$, $\rho_r(z)$ and $\rho_G(z)$ while for $\rho_s(z)$ and $\rho_e(z)$ it is negative. Moreover, the order of the rescaled velocity $\langle v \rangle/v_0$ plateaus in Fig. \ref{fig:3} (a) from the largest to the smallest is the same as the sequence of the probability difference $\Delta P$ for $\zeta = 3.6$ in Fig. \ref{fig:7}. 

The magnitude of $\langle v \rangle/v_0$ plateau can be calculated from Eq. (\ref{eq:R_zeta}). For instance, for the deterministic distribution $\rho_d(z)$ we obtain the closed analytical expression
\begin{equation}
	\frac{\langle v \rangle}{v_0} = 1 + \frac{L - \zeta}{\zeta} = \frac{L}{\zeta}
\end{equation}
which for $L = 2\pi$ and $\zeta = 3.6$ yields $\langle v \rangle/v_0 = 1.74$ as it is shown in Fig. \ref{fig:3} (a).  It follows that in the regime of rare spikes $\lambda \to 0$ the origin of free particle transport boost lies in a profit from the spatial period $L$ of the potential $U(x)$ rather than its barrier height $\varepsilon$. It is also confirmed by the fact that in such regime an increase of $\varepsilon$ in \mbox{Fig. \ref{fig:3}} does not change the magnitude of the rescaled velocity $\langle v \rangle/v_0$.

In Fig. \ref{fig:3} (b) we can notice that the free particle transport boost in the limit of rare spikes $\lambda \to 0$ may be much greater than the upper bound $\langle v \rangle/v_0 = 2$ for the deterministic $\rho_d(z)$ amplitude distribution. It is so for the skew-normal statistics $\rho_s(z)$ for which the plateau reads $\langle v \rangle/v_0 \approx 12$. Since the average distance of the particle relaxation is bounded from above $\langle \Delta x_R \rangle \le \zeta_c = L/2$ it can happen only for smaller mean amplitude $\zeta$. However, when $\zeta < \zeta_c$ amplitude distributions that allow for only positive $\delta$-spikes possess negative balance $\Delta P < 0$ between the probabilities for beneficial and detrimental jump-relaxation processes, c.f. Fig. \ref{fig:7}. The same is true also for the bidirectional and symmetric statistics such as the Gaussian distribution $\rho_G(z)$. The giant free particle transport boost $\langle v \rangle/v_0$ is therefore possible exclusively for the bidirectional and asymmetric amplitude distributions like the skew-normal $\rho_s(z)$ density. It is worth to note also the role of variance $\sigma^2$ independence. If $\zeta \to 0$ the potential barrier crossing events are possible when $\sigma^2$ is not a function of $\zeta$. Therefore only the Gaussian $\rho_G(z)$ and skew-normal $\rho_s(z)$ statistics lead to the non-zero rescaled transport velocity $\langle v \rangle/v_0$ in Fig. \ref{fig:3} (b).

If the constraint of fixed mean bias $\langle \eta(t) \rangle = v_0 = \lambda \zeta = const.$ is imposed, the mean amplitude $\zeta$ must follow $\zeta = v_0/\lambda$. In the regime of rare spikes $\lambda \to 0$ implies $\zeta \to \infty$. Since $\langle \Delta x_R \rangle \le \zeta_c = L/2$ from Eq. (\ref{eq:R_zeta}) it follows that there is no free particle transport boost $\langle v \rangle/v_0 = 1$ regardless of $\delta$-spikes amplitude statistics as it is shown in Fig. \ref{fig:4}.

\subsubsection{Frequent spikes $\langle \tau_P \rangle \ll \langle \tau_R \rangle$}
When the mean waiting time between the successive $\delta$-spikes is much smaller than the average relaxation time $\langle \tau_P \rangle \ll \langle \tau_R \rangle$ the particle is constantly agitated by active fluctuations $\eta(t)$. It implies that the particle does not have enough time to slide down the potential and therefore the relaxation process can be neglected $\langle \Delta x_R \rangle = 0$. The limit of frequent spikes $\langle \tau_P \rangle \ll \langle \tau_R \rangle$ is realized when the frequency $\lambda \to \infty$. One can conclude from the inspection of Fig. \ref{fig:3} and Fig. \ref{fig:4} that in such a case the free particle transport boost does not occur. For amplitude distributions $\rho(z)$ that allow only for positive $\delta$-spikes the rescaled velocity either equals zero or tends to one. The first situation emerges when the mean amplitude of $\delta$-impulse is smaller than the distance between the minimum and maximum of the potential $\zeta < \zeta_c$. The second scenario takes place if $\zeta > \zeta_c$. The exception is the bidirectional amplitude statistics such as the Gaussian $\rho_G(z)$ and the skew-normal distribution $\rho_s(z)$ for which even when $\zeta < \zeta_c$ the rescaled velocity $\langle v \rangle/v_0 = 1$ due to the fact that their variance $\sigma^2$ is a parameter independent of its mean $\zeta$ and therefore the potential barrier crossing events are still possible.

\begin{figure}[t]
    \centering
    \includegraphics[width=1.0\linewidth]{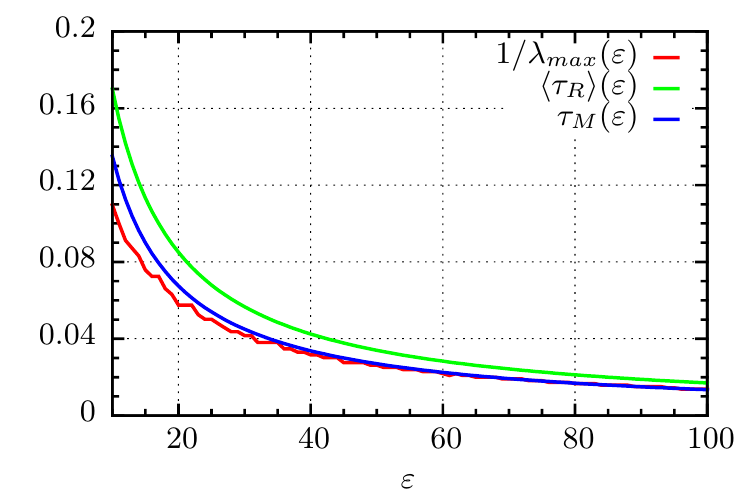}
    \caption{The characteristic time $1/\lambda_{max}$ between the consecutive $\delta$-impulses of active fluctuations $\eta(t)$ corresponding to the rescaled velocity $\langle v \rangle/v_0$ maxima in Fig. \ref{fig:4} for the skew-normal $\rho_s(z)$ amplitude distribution, the average $\langle \tau_R \rangle$ and median $\tau_M$ of the relaxation time all depicted as a function of the potential barrier height $\varepsilon$. The parameters are the same as in Fig. \ref{fig:6}.}
    \label{fig:8}
\end{figure}
\subsubsection{Resonance $\langle \tau_P \rangle \approx \langle \tau_R \rangle$}
As it is often the case in physics, phenomena occurring at the border of two separate physical realms are the most intriguing ones. It is not different this time. When the mean waiting time between $\delta$-spikes and the average relaxation time are matched $\langle \tau_P \rangle \approx \langle \tau_R \rangle$ we detect the resonance behavior in which the giant boost of the free particle transport emerges.

From our toy model of the jump-relaxation process we learned that when the $\delta$-spikes are rare $\langle \tau_P \rangle \gg \langle \tau_R \rangle$ the free particle transport enhancement occurs due to the fact that active fluctuations $\eta(t)$ allow the particle to exploit the spatial periodicity $L$ of the potential $U(x)$ when it relaxes towards the minimum.  If $\delta$-spikes are too frequent $\langle \tau_P \rangle \ll \langle \tau_R \rangle$ the transport boost ceases to exist because the particle is constantly agitated by active noise $\eta(t)$ and it cannot take advantage of the periodic structure $U(x)$. On the other hand, if $\delta$-spikes are rare the particle wastes time in the potential minimum until the next impulse arrives and therefore the transport is not optimal. Since the transition between the rare and frequent $\delta$-spikes is continuous there must exist the corresponding impulse frequency $\lambda_{max}$ for which the transport is optimized in the periodic potential with a given barrier height $\varepsilon$. It translates to a situation when the particle does not waste time in the potential minimum but instead it exploits the relaxation and immediately after that the next $\delta$-spike arrives.

Our claims are confirmed in Fig. \ref{fig:4} where the giant boost of the free particle transport $\langle v \rangle/v_0 \gg 1$ occurs for the skew-normal statistics $\rho_s(z)$ in the resonance case $\langle \tau_P \rangle \approx \langle \tau_R \rangle$. This fact must be contrasted with the rare spikes regime $\langle \tau_P \rangle \gg \langle \tau_R \rangle$ in which free particle transport can be amplified but to a much lesser extent $\langle v \rangle/v_0 > 1$ even for the carefully optimized deterministic amplitude distribution $\rho_d(z)$. Moreover, as we explained above when $\delta$-spikes are scarce in time the free particle transport boost emerges because the particle utilizes the spatial periodicity $L$ of the potential. In contrast, for the resonance regime $\langle \tau_P \rangle \approx \langle \tau_R \rangle$ the particle exploits the potential steepness $\varepsilon$ rather than its period $L$ to achieve transport enhancement. 

When $\varepsilon$ grows the average relaxation time $\langle \tau_R \rangle \propto 1/\varepsilon$ decreases and consequently $\langle \tau_P \rangle = 1/\lambda$ must be reduced to maintain $\langle \tau_P \rangle \approx \langle \tau_R \rangle$. It is indeed the case since if $\varepsilon$ increases the optimal frequency $\lambda_{max}$ for which the free transport boost is maximal gets larger. In Fig. \ref{fig:8} we present comparison of $\lambda_{max}$ with the mean $\langle \tau_R \rangle$ and median $\tau_M$ relaxation time (see the Appendix) all depicted as a function of the potential barrier $\varepsilon$. When the latter increases initial divergence between them quickly dies out and these characteristic times become equivalent in the giant transport regime which proves the resonance-like behavior. However, $1/\lambda_{max}$ is slightly smaller than $\langle \tau_R \rangle$. It follows from the potential $U(x)$ profile for which the particle relaxation is much slower near the minima. It turns out that getting close to them is more optimal than insistently trying to reach them.

Moreover, for $\langle \eta(t) \rangle = v_0 = \lambda \zeta = const.$ the growth of $\lambda$ must be compensated by a decrease of the mean amplitude $\zeta$. As we revealed only the skew-normal $\delta$-spikes statistics $\rho_s(z)$ possesses positive difference $\Delta P > 0$ between the constructive and destructive jump-relaxation processes for vanishing $\zeta \to 0$. This fact explains why in the resonance regime $\langle \tau_P \rangle \approx \langle \tau_R \rangle$ the free particle transport enhancement is observed only for the skew-normal amplitude distribution $\rho_s(z)$. It leads to a  non-trivial and counter-intuitive conclusion that in periodic systems active fluctuations with randomly distributed positive and negative $\delta$-spikes can induce significantly greater directed transport than for active noise of equal average but with only positive amplitudes, even the carefully tuned deterministic ones.


\section{Conclusions}
In this work we thoroughly investigated a paradoxical effect in which a free particle transport induced by active fluctuations in the form of white Poisson shot noise can be boosted by many orders of magnitude when the particle is additionally subjected to a periodic potential. In doing so we focused on the impact of active noise amplitude statistics as well as its spiking rate on the occurrence of this phenomenon. By resorting to a toy model of the jump-relaxation process we identified different regimes of the free particle transport boost and explained in detail their corresponding mechanisms. Our study of various active fluctuations amplitude statistics allowed us to understand and formulate conditions that are necessary for the emergence of giant enhancement of the free particle transport induced by the periodic potential.

In particular, as we revealed the boost of free particle transport can occur in two distinct physical regimes related to the characteristic time scale between two successive $\delta$-spikes of active fluctuations. When $\delta$-impulses are rare the effect emerges due to the fact that the particle with the help of active noise can exploit the spatial periodicity of the potential. On the other hand, when the mean waiting time between $\delta$-spikes and the average relaxation time in the periodic potential are matched the resonance occurs in which the free particle driven by active fluctuations can make use of the potential barrier rather than its period to maximally enhance its transport. The magnitude of boost depends noticeably on the free particle transport velocity. When the latter is already large there is little or no gain by placing the system in the periodic potential. If the directed motion of free particle induced by active noise is slow the velocity can be enormously boosted when the system is additionally subjected to the periodic potential.

However, there are non-trivial constraints on the statistics of $\delta$-spike amplitudes which must be fulfilled for this paradoxical effect to emerge. Our study of selected parameter regimes reveals that for the symmetric periodic potential it occurs in the resonance case for the skew-normal distribution that is bidirectional, i.e. allows for both positive and negative $\delta$-spikes, its variance forms an independent parameter as well as the distribution is asymmetric. The combination of these features generates an unique property which is missing for other considered amplitude statistics. It is a positive difference between the probabilities for the jump-relaxation processes accelerating and slowing down the transport in the periodic potential in the regime when the mean amplitude vanishes. It must be contrasted with other distributions, in particular those with only positive $\delta$-spikes, for which destructive jump-relaxation processes dominate and the free transport enhancement does not emerge.

Our results are relevant not only for microscopic physical systems but also biological ones such as e.g. living cells which are prototype of nonequilibrium system exposed to both thermal and active noise. Therefore our strategy of exploiting periodic potential for giant enhancement of free particle transport driven by active fluctuations may inspire new designs of ultrafast and efficient biologically inspired micro and nanoscale machines. Since we considered a paradigmatic model of nonequilibrium statistical physics that embodies numerous realizations including e.g. a colloidal particle in an optically generated periodic potential \cite{park,paneru} or real biological motors \cite{ezber,ariga} we anticipate stimulating follow-up works of both theoretical and experimental origin. 

\section*{Acknowledgments}
This work has been supported by the Grant NCN No. 2022/45/B/ST3/02619 (J.S.).

\appendix
\section{The relaxation time $\tau_R$}
In the absence of all fluctuations the particle relaxation time towards the potential $U(x)$ minimum is described by the equation
\begin{equation}
	\dot{x} = -U'(x).
\end{equation}
The time $\tau_R(x_A,x_B)$ which the particle needs to move from the point $x_A$ to $x_B$ reads
\begin{equation}
    \tau_R(x_A,x_B)=-\int_{x_A}^{x_B} \frac{dx}{U'(x)}.
\end{equation}
Since the potential $U(x)=\varepsilon \sin{x}$ is periodic we restrict our consideration to the interval $x_A,x_B \in\left(\frac{\pi}{2},\frac{3\pi}{2}\right)$. Note that both the minimum and maximum are excluded because the time required to leave the maximum or reach the minimum is infinite. Then the above formula yields
\begin{equation}
	\tau_R(x_A,x_B)=-\frac{1}{2\epsilon}\ln \left| \frac{1+\sin{x}}{1-\sin{x}}\right|\Biggr|_{x_A}^{x_B}.
    \label{eq_S_sol}
\end{equation}
After the arrival of $\delta$-spike, the particle can jump at any random position $x_A$ and then during the interval $\tau_R(x_A,x_B)$ it is relaxing towards the neighboring potential minimum. The process ends at another random position $x_B$ where the next $\delta$-spike emerges. For this reason it is more adequate to consider the average relaxation time $\langle \tau_R \rangle$ which reads
\begin{equation}
     \langle \tau_R \rangle=\frac{\int_{x_0}^{x_0 + L/2}\int_{x_A}^{x_0 + L/2} \tau_R(x_A,x_B)dx_B dx_A}{\int_{x_0}^{x_0 + L/2}\int_{x_A}^{x_0 + L/2}dx_B dx_A},
\end{equation}
where $x_0$ and $L$ is the potential minimum and its spatial period, respectively. Plugging here Eq. (\ref{eq_S_sol}) we get
\begin{equation}
	 \langle \tau_R \rangle = \frac{1.70}{\varepsilon}.
\end{equation}
Median of the relaxation time $\tau_R(x_A,x_B)$ cannot be calculated in a closed analytic form and therefore must be determined numerically directly from simulations of the underlying deterministic dynamics of the system. The result reads
\begin{equation}
	\tau_M = \frac{1.35}{\varepsilon}.
\end{equation}

\end{document}